\documentclass[conference,twocolumn]{IEEEtran} %IEEEtran article
\usepackage[utf8]{inputenc}
\usepackage[noadjust]{cite}

\usepackage{graphicx}
\usepackage{amsmath}
\usepackage{amssymb}
\usepackage{mathtools}

\usepackage[caption=false,font=footnotesize]{subfig}
\usepackage{nicefrac}
\usepackage{placeins}
\usepackage{threeparttable}
\usepackage{makecell}    
\newcommand{\keywords}{
		Dataset, Internet of Things (IoT), LoRa, low-power wide-area network (LPWAN), propagation model
}
\usepackage[
	pdftex, 
	linktocpage  = false,      		  % ToC, LoF, LoT place hyperlink on page number, rather than entry text
	breaklinks   = true,       		  % so long urls are correctly broken across lines
	hidelinks    = true,       		  % Removes the box around links and makes the active links in \textcolor
	pdftitle={Dataset and UAV Propagation Channel Modeling for LoRa in the 860 MHz ISM Band},      				  % The following is used to annotate the pdf
	pdfauthor={Joachim Tapparel, Andreas Burg}, 
	pdfsubject={}, 
	pdfkeywords={\keywords},
]{hyperref}

\usepackage{tikz} 
\pgfdeclarelayer{background} 
\pgfsetlayers{background,main}
\usepackage{pgfplots}
\usepgfplotslibrary{units}
\usepgfplotslibrary{fillbetween}
\usetikzlibrary{arrows}
\usetikzlibrary{automata,positioning}
\usetikzlibrary{shapes.geometric} 
\usetikzlibrary{shadows.blur}
\usetikzlibrary{shapes.symbols}

\usepgfplotslibrary{external}  
\tikzsetexternalprefix{fig/autogenerate/}

\usepackage[textsize=scriptsize,textwidth=45pt]{todonotes}
\let\OriginalTodo\todo
\RenewDocumentCommand{\todo}{s o m}{%
\tikzexternaldisable 
\IfBooleanTF{#1}{%
\IfValueTF{#2}{\OriginalTodo*[#2]{#3}}{\OriginalTodo*{#3}}%
}{%
    \IfValueTF{#2}{\OriginalTodo[#2]{#3}}{\OriginalTodo{#3}}%
    }%
    \tikzexternalenable
    } 
     
    \let\oldsimiq\sim
\renewcommand{\sim}{{\,\oldsimiq\,}}
\let\oldtablename\tablename
\renewcommand{\tablename}{\mbox{\oldtablename}}

\definecolor{leman}{RGB}{0,167,159} 
\definecolor{commblue}{RGB}{0,96,172}
\definecolor{vertdeau}{RGB}{194,221,176}
\definecolor{perle}{RGB}{202,199,199}
\definecolor{canard}{RGB}{0,116,128}
\definecolor{groseille}{RGB}{181,31,31}
\definecolor{montrose}{RGB}{243,152,105}
\definecolor{zinzolin}{RGB}{92,36,131}
\definecolor{carotte}{RGB}{236,102,8}
\definecolor{chartreuse}{RGB}{200,211,0}
\definecolor{robinEggBlue}{RGB}{10,195,195}
\definecolor{milanoRed}{RGB}{195,5,5}
\definecolor{customGreen}{cmyk}{.88, 0, .96, .1}

\newlength{\belowcaptionskipcustom}
\begin{document} 
%apply predifined settings that directly do et al after 6 authors
% note that BSTCustom:BSTcontrol should be defined in your .bib file
\bstctlcite{BSTCustom:BSTcontrol}
  
\author{\IEEEauthorblockN{ Joachim Tapparel\IEEEauthorrefmark{1} and Andreas Burg\IEEEauthorrefmark{1}}
	\normalsize\IEEEauthorblockA{\IEEEauthorrefmark{1}Telecommunications Circuits Laboratory, \'Ecole Polytechnique F\'ed\'erale de Lausanne, Switzerland
	}
} 

\title{Dataset and UAV Propagation Channel Modeling\\ for LoRa in the 860 MHz ISM Band}
\maketitle
% \thispagestyle{plain}
% \pagestyle{plain}
% \tikzexternaldisable
% \begin{tikzpicture}[remember picture, overlay]
%   \node[anchor=north east, xshift=-.5cm, yshift=-.5cm] at (current page.north east) {
%     \textcolor{red}{\LARGE\textbf{Deadline: \deadline,\daystodeadline{} days left}}
%   };
% \end{tikzpicture}
% \tikzexternalenable
\begin{abstract}
LoRa is one of the most widely used low-power wide-area network technology for the Internet of Things.
To achieve long-range communication with low power consumption at a low cost, LoRa uses a chirp spread spectrum modulation and transmits in the sub-GHz unlicensed industrial, scientific, and medical~(ISM) frequency bands.
Due to the rapid densification of IoT networks, it is crucial to obtain tailored channel models to evaluate the performance of LoRa networks.
While channel models for cellular technologies have been investigated extensively, specific characteristics of LoRa transmissions operating at long range with a rather small (${\boldsymbol{\oldsimiq} }$\,250\,kHz) bandwidth require dedicated measurement campaigns and modeling efforts. 
In this work, we leverage an SDR-based testbed to gather and publish a dataset of LoRa frames transmitted in a campus environment.
The dataset includes IQ samples of the received frames at multiple locations and allows for the evaluation of channel variations with high time resolution.
Using the gathered data, we derive empirical propagation channel models for LoRa that include receiver correlation over distance for three scenarios: unmanned aerial vehicle~(UAV) line-of-sight~(LoS), UAV non-LoS, and pedestrian non-LoS.
Furthermore, the dataset is annotated with synchronization information, enabling the evaluation of receiver algorithms using experimental data.
\end{abstract} 

\section{Introduction}
In recent years, low-power wide-area networks~(LPWANs) have gained popularity for Internet of Things~(IoT) applications.
With currently more than $41$\,\% LPWAN market share (excluding China, which stands out with an exceptionally high NB-IoT deployment), LoRa~\cite{loraPatent} is one of the most widely used LPWAN technologies~\cite{lpwan_market}.
LoRa uses a chirp spread spectrum~(CSS) modulation to achieve long-range communication with low power consumption and operates in the unlicensed industrial, scientific, and medical~(ISM) frequency bands to reduce operation costs~\cite{loraPatent}.
While the LoRa modulation achieves high sensitivity (up to $-141$\,dBm~\cite{sx1303_datasheet}), the long time-on-air and minimization of the transmission power required for good energy efficiency render LoRa transmissions highly sensitive to channel variations.
LoRa is also an attractive standard for the telemetry of unmanned aerial vehicles~(UAVs) which may move fast and which experience channels that are not obviously identical to those experienced at pedestrian heights. 
Channel modeling for LoRa is also relevant for the study of the performance of LoRa networks, particularly when using network simulation tools to evaluate the impact of interference, coverage, or capacity.
In this work, we gather a dataset of LoRa frames transmitted in a campus environment and model channel statistics over distance for three scenarios: pedestrian non-line-of-sight (NLoS), UAV LoS, and UAV NLoS.
Unlike previous channel modeling works for IoT~\cite{petajajarvi2015coverage,masadan2018lora,Rademacher2021pathloss,batalha2022large,Moradbeikie2024rssi}, we gather data using a software-defined radio~(SDR) implementation of LoRa to obtain a full record of IQ samples of valid frames which can be demodulated. 
This data also serves to obtain channel state information directly from the received and synchronized samples. 
Compared to RSSI values (one per frame) that are available from commercial off-the-shelf devices, our dataset provides a much higher temporal resolution and phase information. 
Furthermore, we deployed four distributed receivers to study channel statistics for multiple receiving points, which contributes to the evaluation of concepts such as distributed antenna systems and cloud radio access networks (C-RANs) that take advantage of multiple remote radio heads (RRHs)~\cite{liu2020nephalai,tapparel2024cran}.
The availability of IQ samples enables more than just RSSI-based channel modeling, and can be used to evaluate advanced receiver algorithms, such as interference cancellation, multi-user detection, jamming mitigation, or joint processing.
\paragraph*{Contributions}
First, we build an open-source dataset of LoRa frames that contains raw IQ samples annotated with the necessary information to enable the evaluation of receiver algorithms on field-collected measurements.
Second, we develop and verify a channel model that captures the spatial correlation of the large-scale fading effects over distance for distributed receivers, in both UAV and pedestrian scenarios.

\section{LoRa Technology Overview} 
The physical layer of LoRa is based on a proprietary bit-interleaved coded modulation.
To achieve long-range communication with low power consumption, LoRa support different data rate configurations by adjusting the spreading factor~(SF) and the bandwidth~(BW) of the CSS modulation.
The spreading factor determines the number of chips per symbol as $2^{\text{SF}}$, where SF ranges from $7$ to $12$, with each increment resulting in a $\sim 3$\,dB increase in sensitivity at the cost of close to doubling the duration of the transmission.
The bandwidth can be set to $125$\,kHz, $250$\,kHz, or $500$\,kHz, with higher bandwidths allowing for higher data rates at the cost of reduced sensitivity.
Additionally, LoRa leverages Hamming codes with code rates~(CR) $\in\{\nicefrac{4}{5},\nicefrac{4}{6},\nicefrac{4}{7},\nicefrac{4}{8}\}$ for error correction.
While LoRa devices can operate with a wider set of configurations, e.g., smaller bandwidth and/or spreading factors, we mention here only the configurations used by the currently deployed LoRa medium access control~(MAC) protocol called LoRaWAN\@.
Unlike the LoRa physical layer, LoRaWAN is an open standard defined by the LoRa Alliance that specifies the MAC and network layers for LoRa-based networks~\cite{lora_alliance}.
The network architecture follows a star-of-stars topology where end devices communicate with gateways that forward the received messages to a central network server.
LoRaWAN also supports limited downlink communication, mainly for acknowledgments and data rate control messages.

\section{Measurement Setup and Dataset}
\begin{figure}
  \centering
  \includegraphics{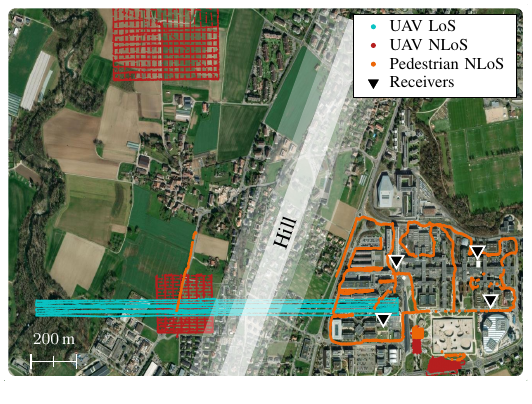}
  \vspace*{-25pt}
  \caption{Map of measurement locations with receivers position}
  \label{fig:measurement_map}
  \vspace*{\belowcaptionskipcustom}
\end{figure}

The measurements were carried out using an SDR-based testbed deployed on the EPFL campus in Lausanne, Switzerland.
The testbed is composed of four gateways (receivers) distributed across roofs of the campus and a mobile LoRa transmitter.
During the measurements, the positions of the receivers are fixed according to \figurename~\ref{fig:measurement_map} and the transmitter is either mounted on a UAV or is carried by a pedestrian.

%------------------------------- Config Transmitter -------------------------------
\begin{table}[t]
  \centering
\begin{threeparttable}[t]
  \renewcommand{\arraystretch}{1.3}
  \caption{Transmitter Configurations }
  \footnotesize
  \label{tab:tx_param}
  \begin{tabular}{l@{\hspace{8pt}} c|l@{\hspace{8pt}} c}
    \hline
    Frequency & $862.5$\,MHz &  Spreading Factor & $7 \,|\, \underline{10}$ \\
    \hline
    Code Rate & $\nicefrac{4}{5}$ & Bandwidth & $125\,|\, \underline{250}$\,kHz \\
    \hline
    Tx Power & $14$\,dBm & Payload Length & $7 \,|\,\underline{19}$\,B  \\
    \hline
  \end{tabular}
  \begin{tablenotes}
    \item[] \textbf{Note:} Underlined values indicate the parameters used for the UAV-mounted transmitter.
    \end{tablenotes}
\end{threeparttable}
\vspace*{\belowcaptionskipcustom}
\end{table}
%-------------------------------------------------------------------------------
\subsection{Gateway}
\begin{figure}
  \centering
  \begin{minipage}{0.48\linewidth}
    \includegraphics{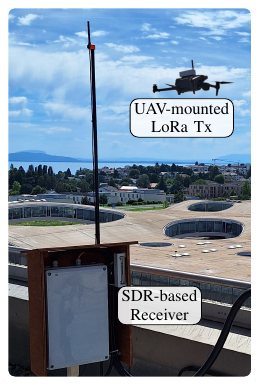}
  \end{minipage}
  \vspace*{-10pt}
  \label{fig:drone_rrh}
  \begin{minipage}{0.48\linewidth}
    \hspace*{-80pt}
    \begin{tikzpicture}
      \node[anchor=south, inner sep=0] (image) at (0,0) {\includegraphics[width=\linewidth]{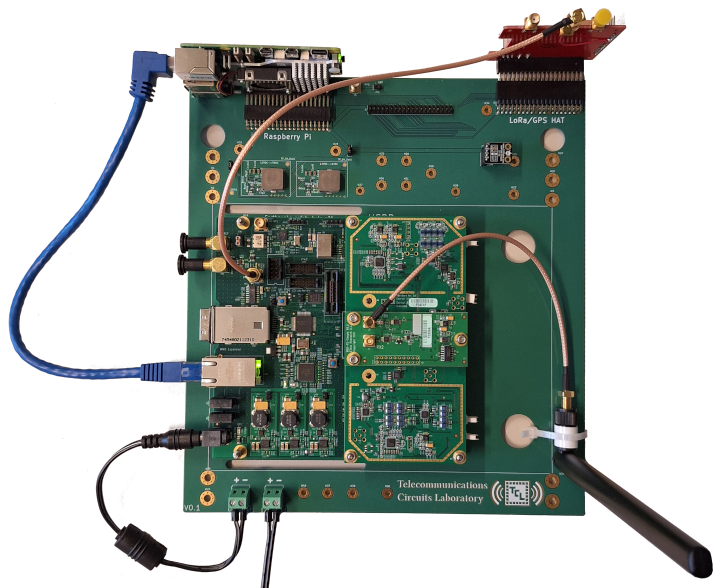}};
      \begin{scope}
        \draw[groseille, very thick, rounded corners] (image.south west) rectangle ($(image.north east)+(0,0.2)$);
        \draw[groseille,densely dashed, very thick] (image.south east)++(-0.2,0) -- ++(-6.7, -0.7);
        \draw[groseille,densely dashed, very thick] ($(image.north west)+(0,0.2)$)++(0.09,-0.01) -- ++(-2.8, -2.75);
        \node[draw=black,anchor=west,fill=white,text opacity=1,inner sep=2pt,fill opacity=0.87,rounded corners,xshift=30pt,yshift=10pt] at (image.west) {\footnotesize NI USRP-2920};
        \node[draw=black,anchor=north west,fill=white,text opacity=1,inner sep=2pt,fill opacity=0.87,rounded corners,xshift=17pt,yshift=4pt] at (image.north west) {\footnotesize Raspberry Pi 5};
        \node[draw=black,anchor=north east,fill=white,text opacity=1,inner sep=2pt,fill opacity=0.87,rounded corners,xshift=-6pt,yshift=-12pt,align=center] at (image.north east) {\footnotesize Quectel \\ \footnotesize L80 GPS};
      \end{scope}
    \end{tikzpicture}
    \vspace*{-50pt}
  \end{minipage}
  \vspace*{12pt}
  \caption{Illustration of the UAV-mounted LoRa transmitter and one of the SDR-based gateway}
  \label{fig:setup_picture}
  \vspace*{\belowcaptionskipcustom}
\end{figure}
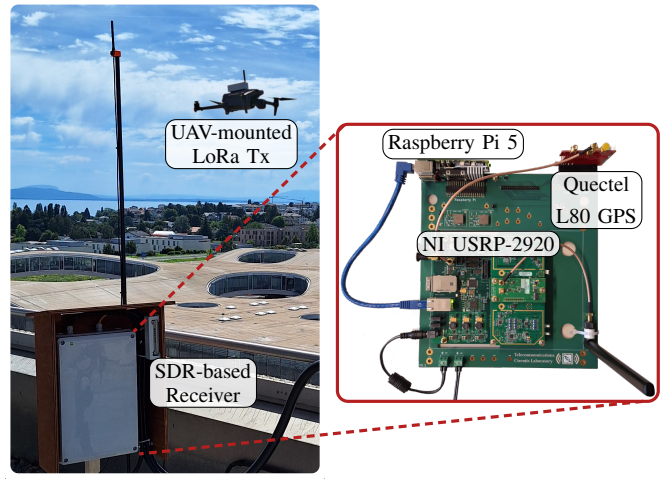

Each gateway is composed of a Raspberry Pi 5, an NI USRP-2920 SDR, and a Quectel L80 GPS receiver.
The SDR implementation is based on the LoRa receiver available in~\cite{lora_sdr_repo} to detect frames, extract the detection information, and store the corresponding IQ samples.
To provide an accurate timestamping of the received samples in the gathered dataset, we use the reference pulse-per-second of the GPS to mitigate the internal clock drift of the USRP\@.
\figurename~\ref{fig:setup_picture} shows one of the four gateways used for the measurements.
%------------------------------- Pictures of RRH -------------------------------
%-------------------------------------------------------------------------------
\subsection{Mobile Transmitter}
Two different mobile transmitters are used for the measurements: a pedestrian-carried LoRa transmitter a UAV-mounted LoRa transmitter.
The pedestrian transmitter is based on a Raspberry Pi 5 connected to a Semtech SX1276 LoRa module and to a Quectel L80 GPS receiver.
A lighter transmitter node composed of an HTCC-AB02S Dev-Board, built around an SX1262 LoRa transceiver and an AIR530Z GPS module is used for the UAV measurements.
\par
During the measurements, the transmitter sends periodic LoRa frames with a payload containing a sequence number and, for the UAV, the transmitter position and instantaneous velocity obtained from the GPS receiver.
The transmission parameters used for the measurements are summarized in \tablename~\ref{tab:tx_param}.
We conducted measurements in representative areas within and around the EPFL campus, as shown in \figurename~\ref{fig:measurement_map}.

\subsection{Dataset Description}
For each of the three considered scenarios, we collected over $17'000$ LoRa frames between all four distributed receivers.
Each entry in the dataset contains the detection information extracted by the SDR-based receiver, as well as the raw IQ samples of the received frame.
Using the information provided for each entry, the IQ samples can be synchronized and demodulated using a LoRa receiver implementation. 
The IQ samples are stored in the standard SigMF file format~\cite{sigmf} with corresponding metadata files containing the annotation with frame information.
The exact content of a dataset entry is indicated in \tablename~\ref{tab:dataset_entry} and the dataset is available at~\cite{tapparel2025loraiq}.
\begin{figure*}[!t]
  \subfloat[UAV LoS]{
    \begin{minipage}{0.33\linewidth}
      \includegraphics{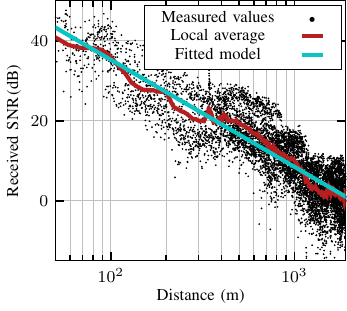}
    \end{minipage}
    \label{drone_los}
  }\hspace*{-6pt}
  \subfloat[UAV NLoS]{
    \begin{minipage}{0.33\linewidth}
      \includegraphics{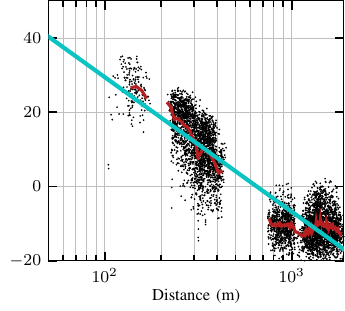}
    \end{minipage}
      \label{drone_nlos}
      }\hspace*{-6pt}
  \subfloat[pedestrian NLoS]{
    \begin{minipage}{0.33\linewidth}
      \includegraphics{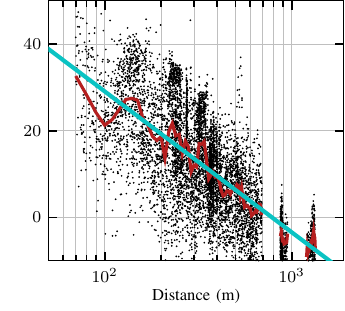}
    \end{minipage}
      \label{ped_nlos}
  }
  \caption{Least square fitting of log-distance model for UAV and pedestrian scenarios}
  \label{fig:pl_model}
\end{figure*}
\begin{table}[!t]
  \renewcommand{\arraystretch}{1.3}
  \caption{Content of one Entry of the Dataset}\label{tab:dataset_entry}
  \centering
  \begin{tabular}{|c|c|c|c|}\hline
    Spreading factor&
    Code rate&
    Carrier frequency&
    Bandwidth\\\hline
    Receiver ID&
    Sampling rate&
    Reception time&
    SNR
    \\\hline
    Tx Position&
    Velocity&
    Carrier freq. offset&
    Frame ID
    \\\hline
    Payload&
    IQ file ID&
    Frame\;start\;in\;file&
    IQ file time
    \\\hline
  \end{tabular}
  \vspace*{\belowcaptionskipcustom}
\end{table}

\section{Channel Modeling}
We first consider a standard log-distance pathloss model and estimate the model parameters for the three considered scenarios.
We then evaluate the large-scale fading effect for each of the distributed receivers and introduce a model for the fading correlation over distance.

\subsection{Log-Distance Channel Model Fitting}
We start by fitting the SNR to a log-distance path loss model
\begin{equation}
  \rho\left(d\right) = \rho_0 - 10 \gamma \log_{10}\left(\frac{d}{d_0}\right) + \mathbf{Z},
\end{equation}
where the impact of fading is described by the random variable~$\mathbf{Z}$ with an unknown distribution to be inferred from the measurements.
We obtain an estimate of the signal-to-noise ratio~(SNR) for each of the received frames using the synchronized IQ samples, and we consider the distance between the transmitter and the receiver at the time of transmission using the GPS coordinates.
\figurename~\ref{fig:pl_model} displays the measured SNR over distance along with the log-distance model with parameters obtained from least-squares fitting for the three aforementioned scenarios.

\begin{figure}[!t]
  \hspace*{-20pt}
  \centering
  \subfloat[UAV LoS]{
    \begin{minipage}{0.49\linewidth}
    \includegraphics{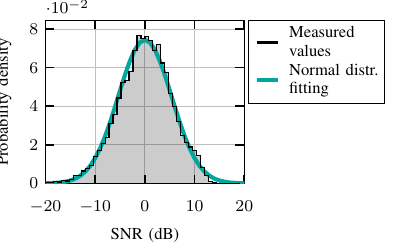}
    \label{drone_los_var}
    \vspace*{-15pt}
    \end{minipage}
    }\\\vspace{-10pt}
    \hspace*{-20pt}
  \subfloat[UAV NLoS]{
    \begin{minipage}{0.49\linewidth}
    \includegraphics{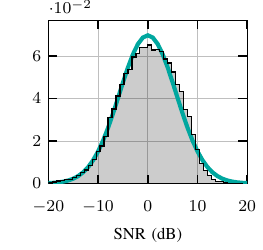}
    \label{drone_nlos_var}
    \vspace*{-15pt}
    \end{minipage}
    }
  \subfloat[Pedestrian NLoS]{
    \begin{minipage}{0.49\linewidth}
    \includegraphics{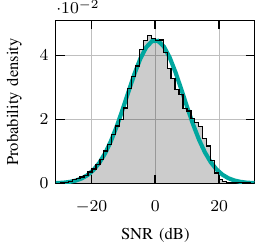}
    \label{ped_nlos_var}
    \vspace*{-15pt}
  \end{minipage}
  }   
\caption{Distribution of the SNR deviation around log-distance model}
\label{fig:var}
\vspace*{\belowcaptionskipcustom}
\end{figure}

\begin{table}[!t]
  \renewcommand{\arraystretch}{1.3}
  \centering
  \caption{Fitting Parameters with a reference distance of $10\,\mathrm{m}$}
\begin{tabular}{l|c|c|c}
  Scenario & $\rho_0$ & $\gamma$ & $\sigma_z$ \\
  \hline
  \hline
  \raisebox{2pt}{\strut}UAV LoS & $64.20$ & $2.79$ &$ 5.36$ \\
  \hline
  UAV NLoS & $65.67$ & $3.62$ &  $5.71$ \\
  \hline
  Pedestrian NLoS & $61.60$ & $3.25 $& $8.90$\\
\end{tabular}
\label{tab:fitting_params}
\vspace*{\belowcaptionskipcustom} 
\end{table}
For all scenarios, the log-distance model closely follows the measured evolution of the average SNR as a function of distance.
The obtained fitting parameters are summarized in \tablename~\ref{tab:fitting_params} using a reference distance $d_0=10$\,m.
The pathloss exponents $\gamma$ obtained from the NLoS measurements are in line with similar evaluations in urban environments~\cite{petajajarvi2015coverage, batalha2022large}.
Additionally, both NLoS scenarios show a pathloss exponent that is close to the value predicted by the Hata-Okumura model~\cite{hata1980empirical} for urban environments ($\gamma_\text{Hata-Oku.}\,{=}\,3.64$) for the considered receiver height of $20$\,m\footnote{The Hata-Okumura model is originally defined for receiver \mbox{heights $\geq 30$\,m}, thus the value $\gamma_\text{Hata-Oku.}$ is obtained by extrapolation.}.
\figurename~\ref{fig:var} shows the distribution of the SNR deviation around the log-distance model for the three considered scenarios.
Each distribution closely follows a Normal distribution.
Therefore, we model the random fading as $\mathbf{Z} \sim \mathcal{N}\left(0,\sigma_z^2\right)$ with standard deviation $\sigma_z$ given in \tablename~\ref{tab:fitting_params}.
While the pathloss exponent for both NLoS scenarios is similar, the standard deviation $\sigma_z$ is significantly higher for the pedestrian NLoS scenario compared to the UAV NLoS scenario, which is consistent with the increased number of scatterers at pedestrian height.

\subsection{Small and Large-Scale Fading with Spatial Correlation}
Although the previous stochastic i.i.d.\ model reproduces well the global SNR evolution with distance, it overlooks the spatial correlation inherent to the large-scale fading experienced by individual receivers. 
As each receiver is influenced by its location and surroundings, a single averaged log-distance model cannot capture these receiver-specific fading patterns.
To model the per-receiver large-scale fading effect, we consider the SNR deviation around the log-distance model for each receiver $r$ separately.
In addition, we separate large-scale and small-scale fading contributions into two independent random variables~$\mathbf{Z}_r \,{=}\, \mathbf{X}_r\,{+}\,\mathbf{Y}_r$.
The small-scale fading~$\mathbf{X}_r$ captures the fast, uncorrelated, variations of the SNR around the large-scale fading effect, while the large-scale fading term~$\mathbf{Y}_r$ captures the slow SNR variations that are correlated over distance.
The SNR measured at receiver $r$ at a distance $d$ is modeled as
\begin{equation}
  \rho_r(d) = \rho_0 - 10 \gamma \log_{10}\left(\frac{d}{10}\right) + \mathbf{X}_r+ \mathbf{Y}_r.
\end{equation}
In the following, we first estimate the small-scale fading component directly from the measurements, without making prior assumptions about the large-scale fading distribution. We then derive a model for the large-scale fading that uses the estimated small-scale fading distribution.
\subsubsection{Small-Scale Fading}
The small-scale fading variable~$\mathbf{X}_r$ captures the fast variations of the SNR around the large-scale fading effect for each receiver $r$.
We obtain an estimate of the small-scale fading values $\mathbf{\hat X}_r$ by subtracting the local average SNR (every $10$ meters) from the measured values.
The small-scale fading component  follows a Normal distribution $\mathcal{N}\left(0,\sigma_{x,r}^2\right)$ for each of the three considered scenarios.
Interestingly, the standard deviation $\sigma_{x,r}$ of the small-scale fading component is similar across all receivers for each scenario (within $1$\,dB from the average).
Therefore, we consider a common standard deviation $\sigma_x$ across all receivers, which is computed using the values $\mathbf{\hat X}_r$ from every receiver.
The resulting parameters for each scenario are summarized in \tablename~\ref{tab:fading_parameters}.
We notice that the standard deviation of the small-scale fading increases the more obstructed the scenario is, which is expected as multipath reflections contribute more significantly to the received signal in NLoS conditions.
\begin{table}
  \renewcommand{\arraystretch}{1.3}
  \centering
  \caption{Fading Parameters}
  \begin{tabular}{l|c|c|c|c}
    Scenario & $\sigma_x$ & $\sigma_y$& $\phi$&$\sigma_\epsilon$  \\
    \hline\hline
    UAV LoS & $2.89$ &$4.51$& $0.974$& $1.02$\\
    \hline
    UAV NLoS & $4.01$&$4.07$& $0.898$& $1.79$\\
    \hline
    Pedestrian NLoS & $7.60$ &$4.63$& $0.750$ &$3.06$\\
  \end{tabular}
  \label{tab:fading_parameters}
  \vspace*{\belowcaptionskipcustom}
\end{table}
\\
\subsubsection{Large-Scale Fading}
As both the general fading $\mathbf{Z}$ and the small-scale fading component $\mathbf{X}_r$ follow two normal distributions, the variance of the large-scale fading $\mathbf{Y}_r$ can be obtained by difference of variances, i.e., $\mathbf{Y}_r\sim \mathcal{N}\left(0,\sigma_y^2\right)$ with $\sigma_y^2\,{=}\,\sigma_z^2\,{-}\,\sigma_x^2$.
Interestingly, the standard deviation $\sigma_y$, indicated in \tablename~\ref{tab:fading_parameters}, is also similar for all three scenarios (within $0.6$\,dB).
This observation shows that, despite the different propagation conditions, the large-scale fading effect remains stable across scenarios, irrespective of whether the transmitter is mounted on a UAV or carried by a pedestrian.
\par
To model the spatial correlation of the large-scale fading component $\mathbf{Y}_r$ for each receiver $r$, we consider an autoregressive model of first order given by
\begin{align}
  \mathbf{Y}_r\left[\left(n+1\right)\Delta_d\right] = \phi\, \mathbf{Y}_r\left[n\Delta_d\right] + \epsilon,
\end{align} 
where $\Delta_d$ denotes the distance step, and where the innovation term and initialization follow  $\epsilon \sim \mathcal{N}\left(0,\sigma_\epsilon^2\right)$ and~$\mathbf{Y}_r\left[0\right] \sim \mathcal{N}\left(0,\sigma_z^2 - \sigma_x^2\right)$.
The autoregressive parameter $\phi$ is estimated using Yule-Walker equations which, for a first order autoregressive process, simplifies to~\cite{kay1981spectrum} 
\begin{equation}
  \hat\phi = \frac{R_{xx}(1)}{R_{xx}(0)},
\end{equation}
where $ R_{xx}(k)$ denotes the autocorrelation function of the experimental large-scale fading values at lag $k$.
The variance of the innovation term $\sigma_\epsilon^2$ is set to ensure that the global variance of the large-scale fading component remains $\sigma_y^2$ as
\begin{equation}
  \sigma_\epsilon^2 = \left(1-\hat\phi^2\right)\left(\sigma_z^2-\sigma_x^2\right).
  \label{eq:innovation_variance}
\end{equation} 
The fading parameters obtained for each of the three scenarios are summarized in \tablename~\ref{tab:fading_parameters} for a reference distance of \mbox{$\Delta_{d_0}\,{=}\,10$\,m.}
We observe that UAV scenarios exhibit a higher spatial correlation (larger $\phi$) and smaller standard deviation $\sigma_\epsilon$ compared to the pedestrian scenario.
Both observations indicate that, while the variance $\sigma_y^2$ is similar across scenarios, the variations occur over larger distances for UAV transmissions.
We note that the reference distance was chosen to provide a good trade-off between capturing the spatial correlation and having enough samples to estimate the large-scale fading autocorrelation function. 
The autoregressive parameter $\phi$ can be adjusted to a different distance $\Delta_d'$ using the relation
\begin{equation}
  \phi' = \phi^{\displaystyle(\nicefrac{\Delta_d'}{\Delta_{d_0}})}.
\end{equation}
The variance of the innovation term $\sigma_\epsilon^2$ can then be adapted by inserting the new value $\phi'$ in~\eqref{eq:innovation_variance}.
\par
\begin{figure}
  \begin{minipage}{\linewidth}
    \subfloat[{UAV LoS scenario}]{
      \hspace*{-23pt}
      \includegraphics{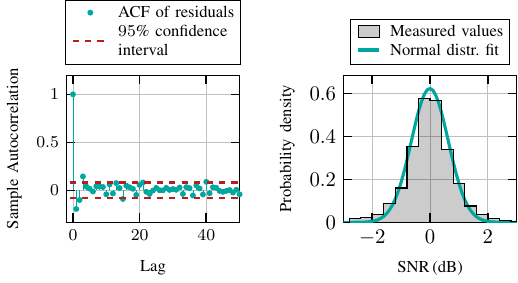}
      \label{fig:residuals_drone_los}
      }
    \end{minipage}
    \vspace*{-2pt}

  \begin{minipage}{\linewidth}
    \subfloat[{UAV NLoS scenario}]{
      \hspace*{-23pt}
      \includegraphics{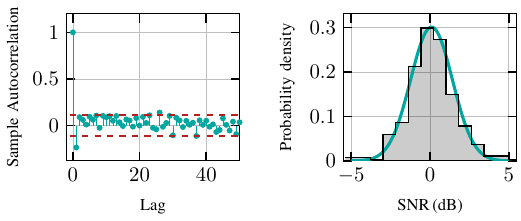}
      \label{fig:residuals_drone_nlos}
    }
  \end{minipage}
  \vspace*{-6pt}
  \begin{minipage}{\linewidth}
    \subfloat[{Pedestrian NLoS scenario}]{
      \hspace*{-23pt}
        \includegraphics{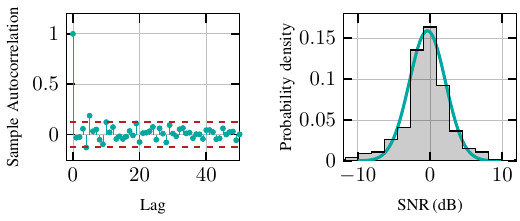}
        \label{fig:residuals_pedestr_nlos}
      }
  \end{minipage}
  \vspace*{3pt}
  \caption{Autocorrelation function and distribution of experimental residuals $\hat\epsilon$ for each scenario}
  \label{fig:residuals_eval}
  \vspace*{\belowcaptionskipcustom}
\end{figure}

To evaluate the fit of the proposed model with the experimentally obtained spatial correlation, we consider the residual error between the measured SNR values and the model prediction.
The residual error is given by
\begin{equation}
  \boldsymbol{\hat{\epsilon}}_r[n] = 
  \mathbf{\tilde Y}_r\left[\left(n+1\right)\Delta_d\right] -\hat\phi\, \mathbf{\tilde Y}_r\left[n\Delta_d\right] ,
\end{equation} 
where $\mathbf{\tilde Y}_r$ denotes the experimental large-scale fading values.
We evaluate both the autocorrelation function and the distribution of the residuals $\boldsymbol{\hat{\epsilon}}
\,{=}\, \big[\, \boldsymbol{\hat{\epsilon}}_{0}^{\top}\;
        \boldsymbol{\hat{\epsilon}}_{1}^{\top}\;
        \boldsymbol{\hat{\epsilon}}_{2}^{\top}\;
        \boldsymbol{\hat{\epsilon}}_{3}^{\top} \,\big]$, as the residuals should be uncorrelated and Gaussian distributed.
As shown in \figurename~\ref{fig:residuals_eval}, most of the autocorrelation values remains within the $95$\,\% confidence bounds of an uncorrelated Gaussian process and each histogram closely follows a normal distribution for each scenario. Together, these results indicate that the proposed large-scale fading model, which assumes Gaussian and white residuals, fits the measured data well.
\subsubsection{Channel Model with Spatial Correlation}
Using the proposed fading model, we can express the SNR values measured by each receiver $r$ over distance as a multivariate normal distribution given by
\begin{equation}
  \boldsymbol{\rho}_\mathrm{r} = \left[\rho_r(0)\; \rho_r(\Delta_d)\; \rho_r(2\Delta_d)\; \dots\right]^\top \sim\;  \mathcal{N}(\boldsymbol{\mu}, \boldsymbol{\Sigma}),
  \end{equation}
  \noindent where $\cdot^\top$ denotes the transpose operation, and where the mean vector and covariance matrix are given componentwise at indices $m,n$ by
  \vspace*{-3pt}
  \begin{align}
  \boldsymbol{\mu}_m &= \rho_0 - 10 \gamma \log_{10}\left( \frac{d}{10} \right), \\[6pt]
  \boldsymbol{\Sigma}_{m,n} &=
    \begin{cases}
    \sigma_z^2, & \text{for } m = n, \\[4pt]
    \dfrac{\sigma_\epsilon^2}{1-\phi^2}\,\phi^{\left|m-n\right|}, & \text{otherwise}.
    \end{cases}
  \end{align}
From the covariance matrix, we can note the main aspects of the proposed model.
First, the variance of the SNR at each distance follows the global measured variance $\sigma_z^2$ and second, the correlation between two SNR values at different distances decays exponentially with the distance as $|\phi| \leq 1$.

\section{Conclusion}
In this article, we presented an SDR-based testbed to evaluate the propagation channel for LoRa transmissions in a campus environment.
Using the testbed, we gathered a dataset of over $75'000$ LoRa frames transmitted in three different scenarios: UAV LoS, UAV NLoS, and pedestrian NLoS.
The public dataset gathered is the first to offer operational IQ sample data from outdoor LoRa transmissions.
We then derived empirical parameter of a log-distance propagation model for LoRa, showing a close match with the measured SNR evolution over distance.
Furthermore, we modeled the spatial correlation of the large-scale fading effect for distributed receivers using an autoregressive model and showed that the proposed model fits the measured data well.
The model parameters that correspond to the different scenarios showed a substantial difference of contribution from small-scale and large-scale fading effects between the UAV and pedestrian scenarios, highlighting the different propagation characteristics experienced at different heights.

\bibliographystyle{IEEEtran}
\bibliography{IEEEabrv,paper.bib}
\end{document}